# Development of a fast plastic scintillation detector with time resolution of less than 10 ps

J.W. Zhao[a], B.H. Sun[a,b,*,1], I. Tanihata[a,b,*], S. Terashima[a,b], L.H. Zhu[a,b], A. Enomoto[c], D. Nagae[d], T. Nishimura[c], S. Omika[c], A. Ozawa[e], Y. Takeuchi[c], T. Yamaguchi[c]

[a] School of Physics and Nuclear Energy Engineering, Beihang University, Beijing, 100191, China

[b] International Research Center for Nuclei and Particles in Cosmos, Beihang University, Beijing, 100191, China

[c] Department of Physics, Saitama University, Saitama 338-8570, Japan

[d] RIKEN Nishina center, Saitama 351-0198, Japan

[e] Institute of Physics, University of Tsukuba, Ibaraki 305-8571, Japan

ABSTRACT

Timing-pick up detectors with excellent timing resolutions are essential in many modern nuclear physics experiments. Aiming to develop a Time-Of-Flight system with precision down to about 10 ps, we have made a systematic study of the timing characteristic of TOF detectors, which consist of several combinations of plastic scintillators and photomultiplier tubes. With the conventional electronics, the best timing resolution of about 5.1 ps ($\sigma$) has been achieved for detectors with an area size of $3\times 1$ cm$^2$. It is found that for data digitalization a combination of TAC and ADC can achieve a better time resolution than currently available TDC. Simultaneously measurements of both time and pulse height are very valuable for correction of time-walk effect.

---

* Corresponding author.
  E-mail address: bhsun@buaa.edu.cn, tanihata@rcnp.osaka-u.ac.jp

## 1. Introduction

The plastic scintillator is probably one of the most widely used organic detectors in nuclear and particle physics. Its mounting, operation and handling are relatively simple, and well-understood, and thus it is often equipped in the Time-Of-Flight (TOF) measurements [1]. A TOF resolution of about 100 ps is enough for most of the general-purpose applications, e.g. for particle identification of radioactive beam with mass number less than 100. However, in recent years there are increasing demands for ultra-fast timing-pick up detectors in high precision measurements. Here we take a TOF mass spectrometry (MS) as an example to illustrate the importance of time determination. The TOF-MS is known as one of the most efficient methods for mapping the nuclear mass surface for short-lived nuclei near the drip line [2]. Currently the single-pass TOF-MS can achieve a mass resolving power of about 5000, and the dominant source is the time resolution of TOF measurements. For a typical flight time of about 500 ns, if the time resolution were improved from current 100 ps to about 10 ps, the final mass resolution of TOF-MS could in principle be enhanced by up to one order of magnitude. This corresponds to an uncertainty of about ±500 keV for the mass of a single nucleus with $A \sim 50$. If succeed, the TOF-MS could significantly speed up the march towards mass measurements of the most neutron-rich nuclei with fairly good precision. The main motivation of present work is to develop a TOF system based on fast plastic scintillators for the TOF-MS measurement of heavy ions at relativistic energies of several hundred MeV/u. We have limited our investigation to plastic scintillators with moderate size that can be used, e.g., at focal planes of fragment separators.

This paper is organized as the followings. The main considerations for fast plastic scintillators with PMT readout are presented in Section 2. After a brief introduction to the experiment in Section 3, we present the results of this work in Section 4. Finally, a summary is given in Section 5.

## 2. General consideration for fast plastic scintillator detector

The timing information of a heavy ion through a plastic scintillation detector is passed on in sequence, from the light yielded during the passage of heavy ions in plastic scintillator, to electrons in PMT, and finally to logic signals from a discriminator and digitalization process. To develop a fast plastic scintillation detector, as already known, the following issues have to be considered and optimized accordingly:

(1) time spread of photons yielded in the scintillator;

(2) dispersion of photons transmitted from scintillator to PMT window;

(3) spectral match between PMT and plastic scintillator;

(4) coupling of plastic scintillators with PMTs;

(5) transit time spread of photoelectrons in the PMT;

(6) time resolution in electronics.

Factor (1) and (5) require using plastic scintillators and PMTs with fast time response, which can be characterized by rise time and decay time in plastic scintillators and short transit time spread in PMTs. Table 1 and Table 2 summarize the features of different types of fast plastic scintillators [3,4] and PMTs [5] used in the present work. Size and thickness of the plastic scintillator will directly affect factor (2). A thick and large scintillator generally enlarges the transit path length and leads to large time spread of photons transmitted to PMT. Concerning factor (3), one needs to match the plastic scintillator with PMT not only in light wavelengths for optimized light collection efficiency but also in the rise times and decay times. As indicated in Table 1 and Table 2, PMTs typically have slower rise time than plastic scintillators. Hence, a very fast plastic and a slow PMT, or vise verse will not necessarily bring any improvement in timing determination, since the poor partner will deteriorate the whole time resolution. The fact that mismatched PMT and plastic scintillator result in worse timing resolution has been seen in previous investigations[6-7].

**Table 1**

Eljen Technology plastic scintillators' characteristics.

| Scintillator | EJ-230 | EJ-232 | EJ-232Q (0.5%) |
| --- | --- | --- | --- |
| Rise time (ns) | 0.5 | 0.35 | 0.11 |
| Decay time (ns) | 1.5 | 1.4 | 0.7 |
| Pulse width, FWMH (ns) | 1.3 | 1.3 | 0.36 |
| Attenuation length* (cm) | 120 | 17 (1) | 6 (1.5) |
| Light output (% Anthracene) | 64 | 55 | 19 |
| Scintillation efficiency, photons/1MeV e- | 9700 | 8400 | ---- |

*Attenuation length of EJ-232 and EJ-232Q (0.5%) are measured in Ref. [7].

**Table 2**

HAMAMATSU PMTs' characteristics.

| PMT | H6533 | H2431 |
| --- | --- | --- |
| Rise time (ns) | 0.7 | 0.7 |
| Transit time (ns) | 10 | 16 |
| Transit time spread (ns) | 0.16 | 0.37 |
| Gain ($\times 10^6$) | 5.7 | 2.5 |
| Effective area size (mm) | Dia. 20 | Dia. 46 |
| Quantum efficiency | ~25% | |

The time resolution $\sigma_T$ of a plastic scintillation detector depends on the statistical fluctuation of the number of photoelectrons $N_{p.e}$ [8-10], and it follows a simple empirical relation with the number of photoelectrons, namely, $\sigma_T \propto (N_{p.e})^{-0.5}$ [6]. A simple way to increase the light collections from plastic to PMT is to increase the number of PMTs for readout. For instance, a widely used concept for bar-like plastic scintillator is to install two PMTs at both ends. The resulting two independent measurements can improve the time resolution by roughly a factor of $\sqrt{2}$ than that with only one PMT as readout. A large size plastic scintillator using up to 32 PMT readouts has been tested, and it was concluded that an intrinsic detector resolution on the order of 10 ps RMS can be obtained [7]. Such configuration is also helpful to minimize the dependence on hit position of heavy ions and can be used directly for position corrections. Moreover, confined beam in space, which is normally carried out by either destructive slits or precise position measurements, is important for TOF measurements with precision down to 10 ps for incident position and/or angle corrections. These additional corrections are especially necessary for thicker detectors. Although less critical, to enhance the photoelectron statistics it is important to take good care of the coupling between plastic scintillator and PMT (i.e. by using light guide and optical coupling materials), and light transmission in plastic scintillator.

Another source affecting time determinations is the intrinsic time resolution of electronics employed. Here the electronics include typically discriminators for time pick up, Time-to-Digital Converters (TDCs), and/or Time-to-Analog Converters (TACs) and Analog-to-Digital Converters (ADCs). In reality, signal dividers and fan-in/fan-out are often used as well. These various electronics contribute to typically 10-30 picoseconds to the time resolution of a plastic scintillation detector. Therefore it is crucial to reduce the electronics modules as much as possible for a fast timing purpose. Another goal of this work is thus to find out the time precision that can be achieved based on the available conventional electronics.

As discussed above, there are many factors that affect the timing properties of a plastic scintillation detector. Quite often we have to find a compromise among these factors. For instance, better light reflection materials such as aluminum foil can help to increase the light collection and thus the number of photoelectrons. However, this will also result in a worse resolution due to the time fluctuation induced by the diffusion in light path lengths. In this work, we use black papers to minimize the reflected lights since the energy losses in detectors are very large.

## 3. Experiment

The experiment was performed at the secondary beam line, SB2 course [11] in Heavy Ion Medical Accelerator in Chiba (HIMAC) at National Institute of Radiological Science (NIRS), Japan. Primary beam of $^{56}$Fe at 500 MeV/u and secondary beam at about 200MeV/u were used. A schematic drawing of our detector arrangement is shown in Fig.1. Two plastic scintillation counters (PL1 and PL3) at the dispersive focus F1 and the achromatic focus F3 were used to measure the TOF. Two silicon counters (Si1 and Si2) installed at F2 were used to measure the energy deposit (dE) of particles of interest. A ϕ16 mm collimator was placed right before the PL3 counter. Our detectors ($S_1$-$S_6$) were placed in the downstream of PL3.

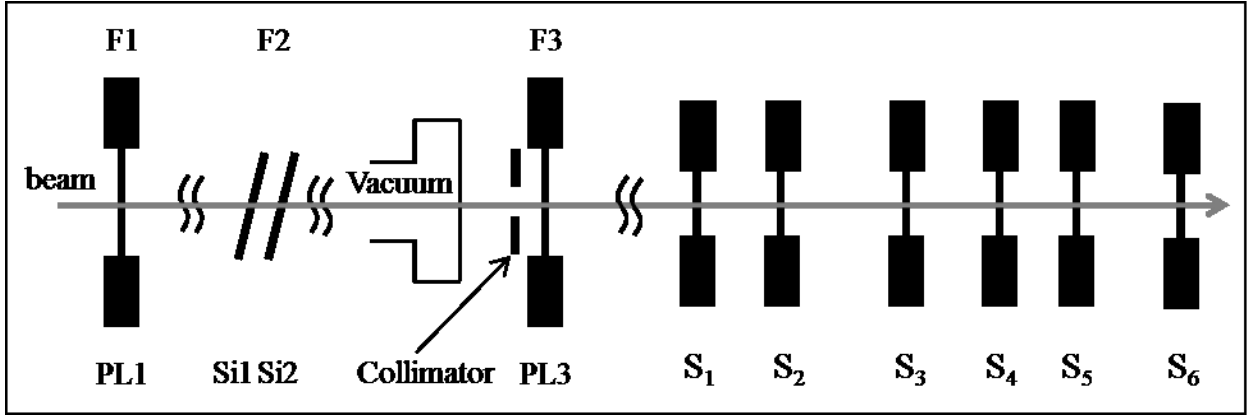

Fig. 1 Detector arrangement in the present experiment. The plastic scintillation counters PL1 at focal plane F1 and PL3 at F3, and two silicon detectors at F2, were used to measure the TOF and energy loss of the heavy ions, respectively. Detectors, $S_1$-$S_6$, were placed in the downstream of PL3.

PMTs were connected to both ends of the plastic scintillators through optical silicone rubber EJ-560 to improve the light transmission. We used several types of fast plastic scintillators and fast PMTs to construct six detectors for a systematic study. The plastic scintillators used here include EJ-230, EJ-232 and EJ-232Q made by ELJEN Technology. EJ-232Q (with 0.5% quenching level of benzophenone) is a quenched version of EJ-232. The characteristics of the plastic scintillators are shown in Table 1. The sizes of all scintillators are 30×10 mm$^2$, except that an extra EJ-232Q is 50×50 mm$^2$. The thicknesses of the plastic scintillators are all fixed to 3mm. The PMTs used here include types of H6533 and H2431 made by HAMAMATSU Company. Characteristics of the PMTs are listed in Table 2.

The schematic diagram of electronics arrangement is shown in Fig. 2. Signals generated by the plastic scintillation detectors are split into two to provide both energy and time information. One is delivered to a Charge-to-Digital Converter (QDC) for energy loss measurement. The other is fed to a leading edge discriminator (LED). Discriminator thresholds are set as low as possible, but above the noise level of PMTs. Considering the long transmission distance (corresponding to a time delay of about 150 ns) between the experimental setup and the DAQ system, we used a second discriminator after the long cable transmission for each timing line to reshape the timing signals.

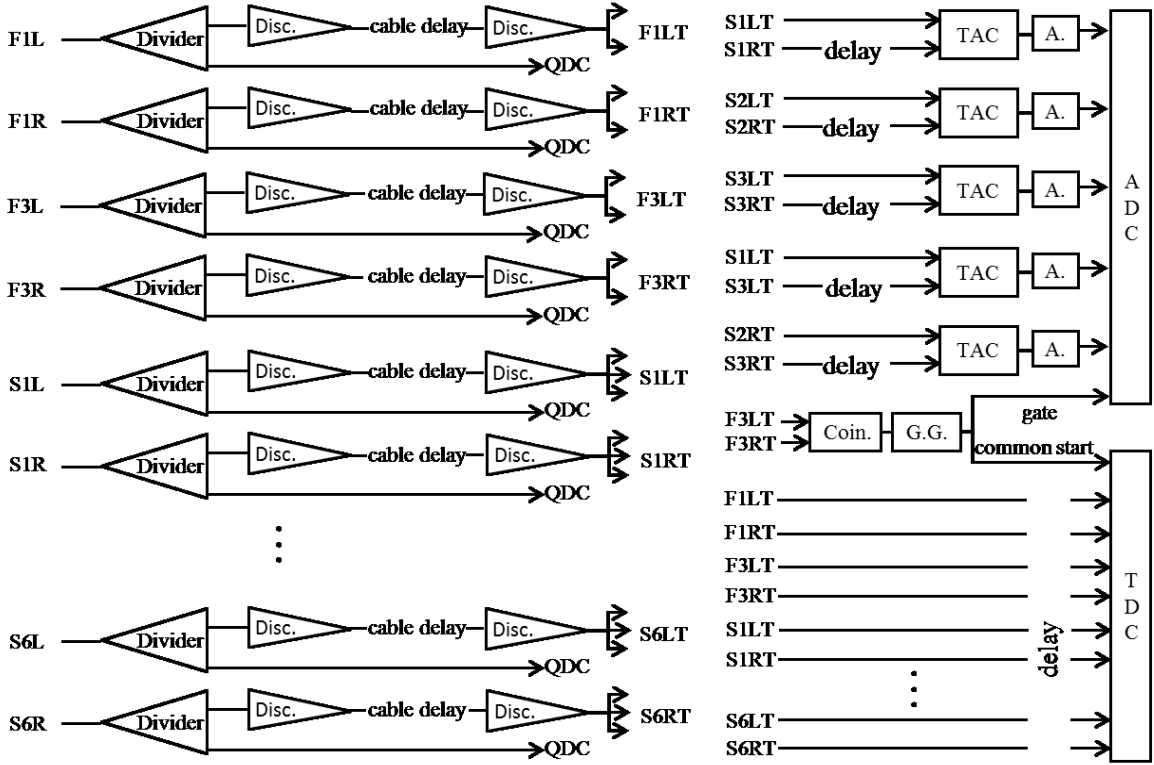

Fig. 2 Schematic diagram of electronics arrangement. F1L (R) is the signal from Left (Right) PMT of TOF-start plastic scintillation counter at F1. F3L (R) is the signal from Left (Right) PMT of TOF-stop plastic scintillation counter at F3. S1L (R) to S6L (R) is the signal from Left (Right) PMT of plastic scintillation detector $S_1$ to $S_6$ respectively. For more details, refer to the text.

The timing information is recorded with both time-to-digital converter (TDC) and time-to-analog converter (TAC) for comparisons. The output analog signals of TACs are then fed to analog-to-digital converter (ADC). As shown in next section, a combination of TAC and ADC (TAC+ADC) can in principle achieve a better time resolution than currently available TDC. In our experiment pulse height attenuators (represented with "A." in Fig. 2) have to be applied between TAC and ADC to match the range of ADC. The modes of LEDs, TACs, TDC, QDC and ADC used here are LeCroy 623B, ORTEC 567, Phillips 7186, REPIC RPC-022 and HOSHIN C008, respectively.

## 4. Data analysis and results

The timing of each detector is calculated by taking the average time from both PMTs at left and right side of a plastic scintillator. For example, the timing $T_1^{raw}$ of detector $S_1$ is

$$T_1^{raw} = \frac{T_{1R}+T_{1L}}{2}, \qquad (1)$$

where $T_{1R}$, $T_{1L}$ is the time determined from the right and left side PMT, respectively. As mentioned in previous section, such average can minimize the hit position uncertainty. Furthermore, in such method we find no significant dependence of time resolution on the position. The position of each ion can be determined by two PPAC detectors placed before the plastic scintillators $S_1$-$S_6$.

To correct the time walk due to the variance of pulse heights, we introduce the following formula:

$$T = T^{raw} - \frac{C_w}{\sqrt[4]{Q_R \times Q_L}}. \quad (2)$$

Here, $T^{raw}$ and $T$ are the measured time and corrected time with pulse height. $Q_R$ and $Q_L$ are the pulse heights of right and left side PMT. $C_w$ is the correction coefficient to be determined. Similar method has been used in previous works[6,12-14].

The TOF between various detectors, e.g. $S_1$ and $S_3$, $T_{31}$ is thus

$$T_{31} = T_3 - T_1. \quad (3)$$

In Fig. 3(a) there is a weak dependence of TOFs on the pulse heights due to the walk effect and such effect can be well corrected by Eq. (2) as shown in Fig. 3(b). Moreover, the TOF distribution with walk effect correction has about 20% of improvement in time resolution. It is found that the TOF distribution can be described very well by a Gaussian function. We thus use the Gaussian parameter σ (standard deviation) to evaluate the time resolution hereafter. By comparing the TOF distribution in Fig. 3(c), one can see that the time-walk correction is critical for high precision time determination and it can improve the time resolution by about 20%, from 14.6ps to 11.6 ps. Such walk effect is also seen in all the other detectors.

Assuming that the timing distribution of each detector follows Gaussian distribution, the time resolutions $\sigma_i$ of an individual detector $S_i$ can be determined by using any three detectors as a system. For example, for detectors $S_1$, $S_2$ and $S_3$, we can have three linear relations, i.e.

$$\sigma_{21}^2 = \sigma_2^2 + \sigma_1^2, \quad (4)$$

$$\sigma_{31}^2 = \sigma_3^2 + \sigma_1^2, \quad (5)$$

$$\sigma_{32}^2 = \sigma_3^2 + \sigma_2^2. \quad (6)$$

Here $\sigma_{ij}$ represents the resolution of the TOF between detector $S_i$ and detector $S_j$.

The individual time resolution $\sigma_i$ can therefore be determined from three measurements of $\sigma_{ij}$. It should be noticed that the time resolution here contains all the source of contributions as discussed in Section 2. In the following discussions, we will concentrate on the dependences of time resolutions on bias voltages applied to PMTs, types of plastic scintillators and PMTs, and also signal digitizing methods.

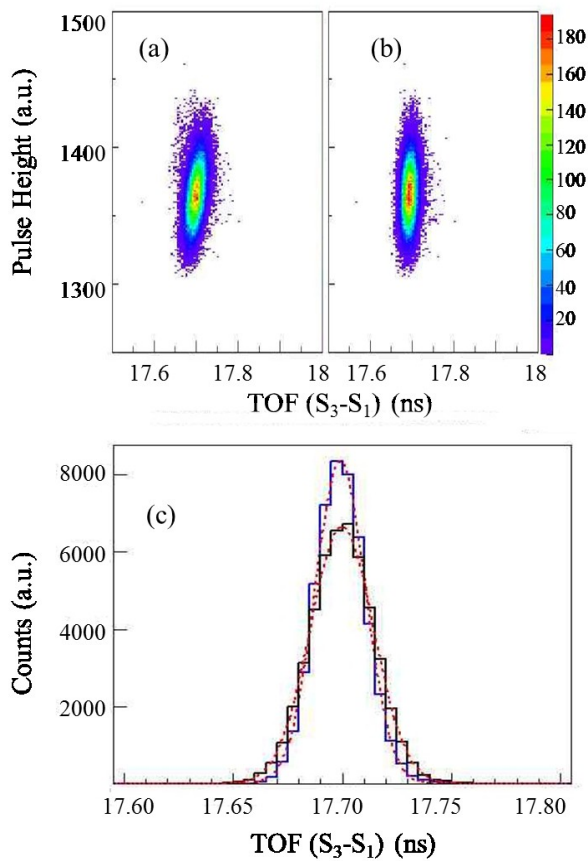

Fig. 3 Correlation between TOF and pulse height of one detector, (a) without and (b) with time-walk correction. The corresponding TOF distribution of (a) (black solid line) and (b) (blue solid line) are shown in panel (c). Gaussian fittings of the TOF distributions are also shown (red dash line).

*4.1 bias voltages applied to PMTs*

A higher voltage applied to PMT is generally preferred for fast timing purpose because transit time spread in PMT is smaller. However, in reality the voltages usually need to be carefully adjusted to avoid too large current in PMT. This is in particular the case for heavy ion detections. For instance, in the present experiment the energy loss in 3 mm thick plastic amounts to be more than 500 MeV.

Figure 4 shows that the time resolution of each detector is not very sensitive to the variance of the applied bias voltage. This is due to the very large energy deposit in the scintillators. Although not tested systematically in present experiment, most likely one can F operate the PMTs at even lower bias voltage and meanwhile keep the good time precision. This result agrees well with previous investigations for heavy ions [6,15]. In the following studies, we have fixed the bias voltage to -1700, -1350 and -1400 for detectors $S_1$, $S_2$ and $S_3$, respectively.

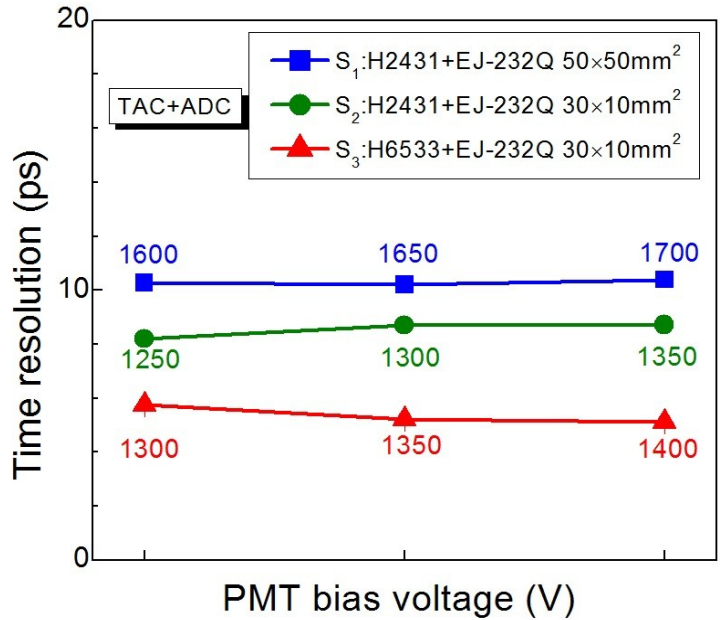

Fig. 4 Time resolutions ($\sigma$) obtained for three different detectors with PMTs operated at various voltages. For instance, Detector $S_1$ stands for a detector composed of a PMT of H2431 type and a $50 \times 50$ mm$^2$ plastic scintillator of EJ-232Q type. The bias voltages applied are indicated along with the relevant symbols. Timing signals are recorded with TAC+ADC.

*4.2 PMT and scintillator*

We select two types of fast PMTs, H2431 and H6533 for comparison. The characteristics of the PMTs are listed in Table 2. H2431 has a much larger effective active area than H6533, but a factor of two worse in transit time spread. The rise time of these two types of PMTs are typically a factor of two more than that of the fast plastic scintillators as shown in Table 1 and 2.

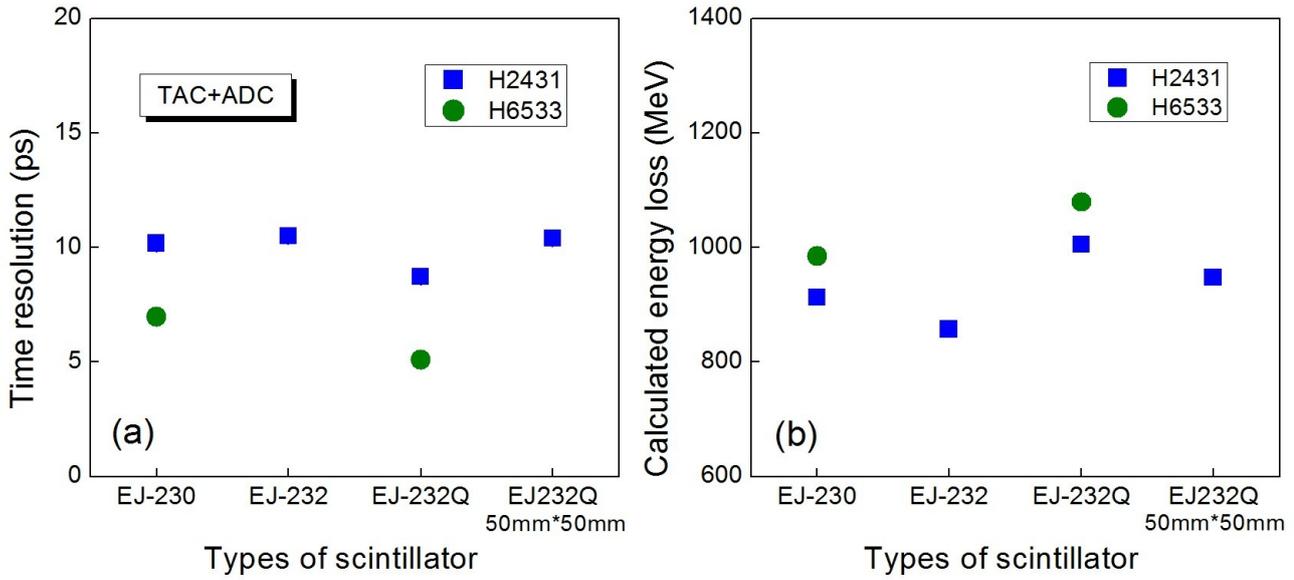

Fig 5. (a) Time resolution (σ) obtained for different combinations of PMTs and scintillators. (b) Calculated energy loss in the relevant scintillator in panel (a). Energy losses in the scintillators are calculated using ATIMA 1.2 LS theory / LISE++ build-in tool Physical Calculator. The size of scintillators here is $30 \times 10 \times 3^t$ mm$^3$, except the one labeled with 50mm*50mm. Timing information is recorded with TAC+ADC.

The time resolution with different types of PMTs and scintillators is summarized in Fig. 5 (a). In this study, the scintillators' size is confined to $30 \times 10 \times 3$ mm$^3$ except one of $50 \times 50 \times 3^t$ mm$^3$. The contact areas between PMT and scintillator are $10 \times 3$ mm$^2$ and $50 \times 3$ mm$^2$ respectively. It is found that a time resolution of better than 11 ps has been achieved for all the different combinations of PMTs and scintillators of interest. For the same plastic scintillator, H6533 gives a better time resolution than H2431. This is consistent with the fact that H6533 has shorter transit time spread. The combination of H6533 and EJ-232Q gives the best time resolution of 5.1 ps in sigma.

For the same PMT, a better time resolution is obtained by using the scintillator EJ-232Q. Moreover, we prepared two EJ-232Q in two different sizes, $30 \times 10 \times 3^t$ mm$^3$ and $50 \times 50 \times 3^t$ mm$^3$. It is evidenced that the gain by using EJ-232Q is fully washed out by an increased size of about a factor of 8. In reality, it is critical to choose a scintillator with fast timing response and meanwhile reasonably small dimension to fit the request.

A worse time resolution is observed for EJ-232 comparing to that of EJ-230. Although EJ-232 has a faster rise time, its lower light output and short attenuation length drag the time resolution down. As indicated in Fig. 5 (b), there exists a clear correlation of energy deposit in detectors with the time resolution.

Moreover, secondary beams of around 200 MeV/u were produced to examine the dependence of time resolution on energy deposit. Five isotopes, $^{54}$Cr, $^{52}$V, $^{51}$V, $^{49}$Ti, $^{48}$Ti with fairly good statistics and clearly separations from neighboring isotopes are used for comparisons. The results are summarized in Fig. 6. Taking detector $S_2$ as an example, the time resolution is improved by about 16% when the energy deposit is increased by 20%. The improvement is even more distinct for detectors S1 and S3. This reflects the fact that the time resolution of the detector strongly depends on the statistics of photoelectrons.

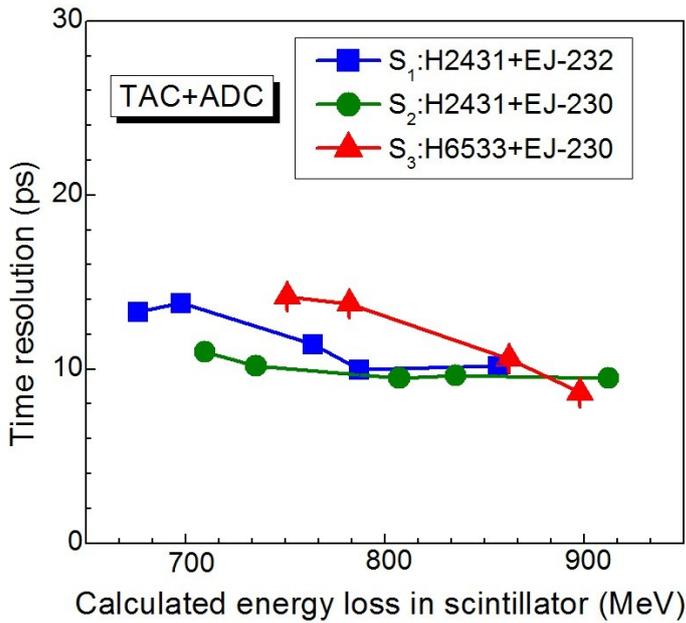

Fig. 6 Time resolution ($\sigma$) vs. energy loss of $^{48}$Ti, $^{49}$Ti, $^{51}$V, $^{52}$V, $^{54}$Cr isotopes in scintillators. Energy losses in the scintillators are calculated using ATIMA 1.2 LS theory / LISE++ build-in tool Physical Calculator. The size of scintillators used here is $30 \times 10 \times 3^t$ mm$^3$. Timing information is recorded with TAC+ADC.

*4.3 Electronics*

To isolate the contribution from electronics, we use a logic fan-in/fan-out NIM signals to replace the detector signals in Fig. 2. In such a way, one can measure the timing jitters between the two identical lines of electronics. Typical time differences between two electronic lines are shown in Fig. 7. Considering 3 ps resolution for the logic fan-in/fan-out, the electronics contribution can thus be determined as 9 ps for each TAC+ADC channel and 22.5 ps from each TDC channel. Consequently, the best intrinsic time resolution down 2.4 ps can been obtained. In other word, the best time precision that one can get from a plastic scintillator is basically governed by the conventional electronics used. A time precision toward 1 ps will need a breakthrough in precision of electronics.

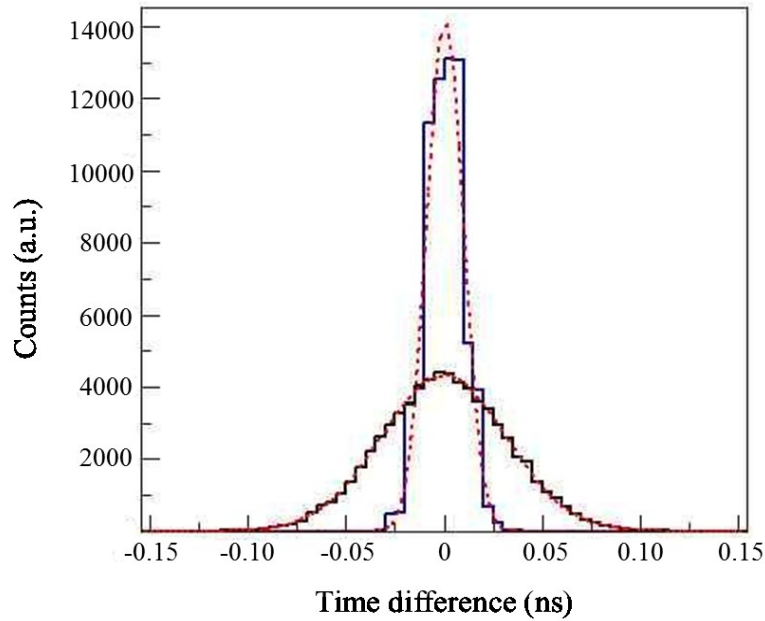

Fig. 7 Time differences between two electronic lines with a logic fan-in/fan-out NIM signals as inputs in Fig. 2. The narrow peak (blue solid line) is a typical distribution of using TAC+ADC, and broad peak (black solid line)is a typical distribution of using TDC. Gaussian fittings of the distributions are also shown (red dash line).

Fig. 8 compares different signal digitizing method, TAC+ADC and TDC. The detectors shown here are the same as those in Fig. 5 (a), but the time were digitized by TDC instead of TAC+ADC. The combination of TAC 567 and ADC of HOSHIN C008 can provide a precision of about 12.5 ps per channel-bin, while TDC of Phillips 7186 is about 25 ps per channel-bin. It is found that the results with TDC are typically about 10 ps worse, but they follow a similar dependence on PMTs and scintillators as TAC+ADC. Recent works aim to develop a high precision compacted TAC+ADC [16-17] may be used for fast timing purpose.

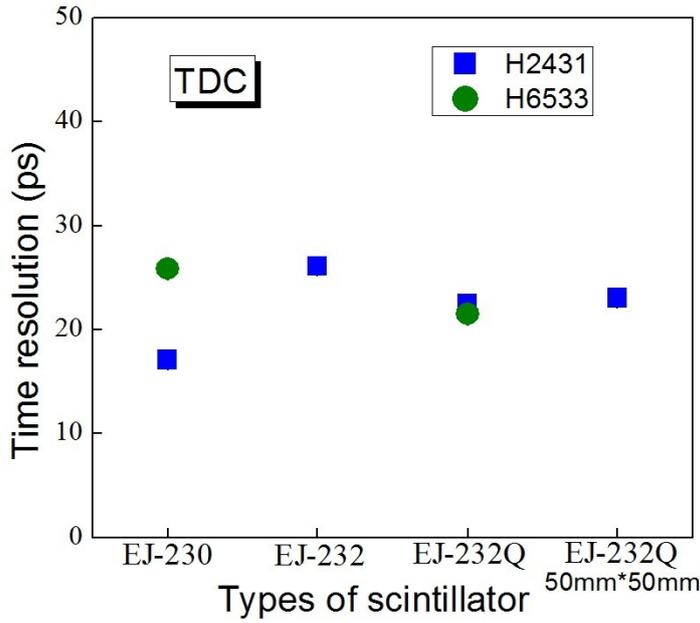

Fig. 8 Same for Fig.5 (a) but recorded with TDC instead of TAC+ADC.

## 5. Conclusion and perspective

In this work, we made a systematic test of plastic scintillator with moderate size using iron beam at relativistic energy of about 500 MeV/u and secondary beam of about 200 MeV/u. The best time resolution down to 5.1 ps, including the contributions from both the detectors and electronics, is obtained with the combination of PMT H6533 and 3mm thick plastic scintillator EJ-232Q (0.5%). No significant dependence on bias voltage applied to PMT was observed in 100 V range. It is found that the energy loss in the scintillator can significantly affect the time resolution, and its measurement is also critical for corrections of time walk, which can improve the time resolution by about 20%. Furthermore, we show that the best time precision is governed by the conversional electronics, which is of typically 10-25 ps. Currently TAC+ADC has an advantage than TDC for signal digitization. For the sake of breaking the developmental "bottle-neck", one need the development in fast signal pickup, sampling, digitization and transmissions, like new electronics and/or novel digitalization techniques, high frequency cables [18].

## Acknowledgements

We appreciate the staff of the HIMAC accelerator for providing a stable beam during the experiment. This work was supported partially by the National Natural Science Foundation of China (Grant Nos. 11235002 and 11475014) and the Program for New Century Excellent Talents in University (Grant No. NCET-09-0031).**Reference**

[1] A. Estradè et al., Physical Review Letters 107 (2011) 172503.

[2] B.H. Sun et al., Frontiers of Physics 10 (2015) 102102.

[3] http://www.eljentechnology.com/index.php/products/plastic-scintillators

[4] A. Ebran et al., Nuclear Instruments and Methods in Physics Research A 728 (2013) 40.

[5] http://www.hamamatsu.com/resources/pdf/etd/PMT_78-85_e.pdf

[6] S. Nishimura et al., Nuclear Instruments and Methods in Physics Research A 510 (2003) 377.

[7] R. Hoischen et al., Nuclear Instruments and Methods in Physics Research A 654 (2011) 354.

[8] W. B. Atwood. SLAC-PUB-2620, 1980.

[9] S. Ahmad et al, Nuclear Instruments and Methods in Physics Research A 330 (1993) 416.

[10] Mizuki Kurata et al., Nuclear Instruments and Methods in Physics Research A 349 (1994) 447.

[11] M. Kanazawa et al., Nuclear Physics A 746 (2004) 393c.

[12] W. Braunschweig et al., Nuclear Instruments and Methods 134 (1976) 261.

[13] T. Tanimori et al., Nuclear Instruments and Methods 216 (1983) 57.

[14] Chong Wu et al., Nuclear Instruments and Methods in Physics Research A 555 (2005) 142.

[15] S. Nakajima et al., Nuclear Instruments and Methods in Physics Research B 266 (2008) 4621.

[16] Karsten Koch et al., IEEE Transactions on Nuclear Science, 52- 3 (2005) 745.

[17] Zule Xu et al., IEEE Transactions on Nuclear Science, 61-2 ( 2014) 852.

[18] W. Zhang et al., Nuclear Instruments and Methods in Physics Research A

756 (2014) 1.